\documentclass[12pt]{article}
\usepackage{amsfonts}
\usepackage{amsmath}
\usepackage{amssymb}
\usepackage{graphicx}
\usepackage{xcolor}
\usepackage[all, knot]{xy}
\usepackage{tikz}
\usepackage{cancel}
\usepackage{comment}

\usepackage{ulem}
\usepackage{hyperref}

\usepackage[utf8]{inputenc}
\usepackage{epstopdf}
\usepackage[footnotesize]{caption}
\usepackage{amsthm}
\usepackage{enumitem}
\usepackage{mathrsfs}

\makeatletter
\newcommand\xleftrightarrow[2][]{%
	\ext@arrow 9999{\longleftrightarrowfill@}{#1}{#2}}
\newcommand\longleftrightarrowfill@{%
	\arrowfill@\leftarrow\relbar\rightarrow}
\makeatother

\usepackage[margin=3cm]{geometry}

\def \be {\begin{equation}}
\def \ee {\end{equation}}
\def \bea {\begin{eqnarray}}
\def \eea {\end{eqnarray}}
\def \nn {\nonumber}

\def \rr {\raise.35ex\hbox{\small $\prime$}\kern-.17em{\mbox{\large $\imath$}}}

\def \dels {\partial\kern-.6em /\kern.1em}
\def \As {{A\kern-.5em / \kern.5em}}
\def \Ds {D\kern-.7em / \kern.5em}

\def \ks {k\kern-.5em /}
\def \ls {l\kern-.5em /}

\def \sgn {\mbox{\small sgn}}

\newcommand{\ci}[1]{}

\newcommand{\ba}{\begin{eqnarray}}
\newcommand{\ea}{\end{eqnarray}}
\newcommand{\bal}{\begin{align}}
\newcommand{\eal}{\end{align}}
\newcommand{\bay}[1]{\left(\begin{array}{#1}}
\newcommand{\eay}{\end{array}\right)}

\newcommand{\hide}[1]{}

\newlist{axioms}{enumerate}{2}
\setlist[axioms,1]{label=\textbf{A\arabic{axiomsi}.}, ref=A\arabic{axiomsi}}
\setlist[axioms,2]{label=\textbf{A\arabic{axiomsi}\rlap{\myEnumCounter{axiomsii}}.},%
                   ref=A\arabic{axiomsi}\myEnumCounter{axiomsii},%
                   align=parleft,%
                   leftmargin=0em,%
                   itemsep=1.4ex,%
                   before={\stepcounter{axiomsi}}}

  \usetikzlibrary{decorations.markings}

\begin{document}

\begin{titlepage}
\begin{center}

\textbf{\LARGE
On the Backreaction of Dirac Matter in\\ 
JT Gravity and SYK Model 
\vskip.3cm
}
\vskip .5in
{\large
Pak Hang Chris Lau$^{a,b}$ \footnote{e-mail address: u948999a@icho2.osaka-u.ac.jp},
Chen-Te Ma$^{c,d}$ \footnote{e-mail address: yefgst@gmail.com}, 
Jeff Murugan$^{e,f}$ \footnote{e-mail address: jeff.murugan@uct.ac.za}, 
and Masaki Tezuka$^g$ \footnote{e-mail address: tezuka@scphys.kyoto-u.ac.jp} 
\\
\vskip 1mm
}
{\sl
$^a$
Department of Physics, Osaka University, Toyonaka, Osaka 560-0043, Japan. 
\\
$^b$
Department of Physics, Kobe University, Kobe-shi 657-8501, Hyogo, Japan. 
\\
$^c$ 
Department of Physics and Astronomy, Iowa State University, Ames, Iowa 50011, US. 
\\
$^d$
Asia Pacific Center for Theoretical Physics,\\
Pohang University of Science and Technology, 
Pohang 37673, Gyeongsangbuk-do, South Korea. 
\\
$^e$
The Laboratory for Quantum Gravity and Strings,\\
Department of Mathematics and Applied Mathematics,
University of Cape Town, Private Bag, Rondebosch 7700, South Africa.
\\
$^f$
National Institute for Theoretical and Computational Sciences, 
Private Bag X1, Matieland,
South Africa.
\\
$^g$ 
Department of Physics, Kyoto University, Kitashirakawa, Sakyo-ku, Kyoto 606-8502, Japan.
}\\
\vskip 1mm
\vspace{40pt}
\end{center}

\newpage
\begin{abstract} 
We model backreaction in AdS$_2$ JT gravity via a proposed boundary dual Sachdev-Ye-Kitaev quantum dot coupled to Dirac fermion matter and study it from the perspective of quantum entanglement and chaos. 
The boundary effective action accounts for the backreaction through a linear coupling of the Dirac fermions to the Gaussian-random two-body Majorana interaction term in the low-energy limit. 
We calculate the time evolution of the entanglement entropy between graviton and Dirac fermion fields for a separable initial state and find that it initially increases and then saturates to a finite value. 
Moreover, in the limit of a large number of fermions, we find a maximally entangled state between the Majorana and Dirac fields in the saturation region, implying a transition of the von Neumann algebra of observables from type I to type II. 
This transition in turn indicates a loss of information in the holographically dual emergent spacetime. 
We corroborate these observations with a detailed numerical computation of the averaged nearest-neighbor gap ratio of the boundary spectrum and provide a useful complement to quantum entanglement studies of holography.
\end{abstract}
\end{titlepage}

\section{Introduction}
\label{sec:1}
\noindent 
Einstein gravity is, famously, trivial in 2-dimensional spacetime. 
To circumvent this, one can couple the graviton to a scalar, the dilaton \cite{Teitelboim:1983ux,Jackiw:1984je}. 
The resulting Jackiw-Teitelboim (JT) gravity furnishes a simple, yet powerful, laboratory in which to study various aspects of quantum gravity, without the complexities that plague gravity in higher dimensions. 
Popular in the mid 80's, JT gravity has seen a revival of interest over the past decade. 
This recent surge in activity is a result of several factors:
\begin{itemize}
    \item 
    JT gravity's exact solvability in (1+1) dimensions offers a unique analytical study of the boundary theory and its backreaction. 
    This simplicity stands in stark contrast to higher-dimensional gravity, which grapples with non-renormalizability and unruly quantum corrections. 
    JT gravity's simplicity and controllability make it an ideal testing ground for Quantum Gravity principles, encompassing black hole physics, holography, and the Anti-de Sitter/Conformal Field Theory (AdS/CFT) correspondence.
    \item 
    JT gravity has a boundary description called the Schwarzian theory \cite{Jensen:2016pah,Maldacena:2016upp,Engelsoy:2016xyb}, which it shares with the low energy limit of a remarkable UV-complete, 1-dimensional quantum mechanical system called the Sachdev-Ye-Kitaev (SYK) model \cite{Sarosi:2017ykf,Chowdhury:2021qpy}. 
    The latter describes the quantum mechanics of $N$ Majorana fermions with $q$-Fermi all-to-all random interactions and is known for its non-trivial behavior in the large-$N$ and low-energy limits, where it exhibits an emergent conformal symmetry and a Schwarzian effective action.
    \item 
    The simplifications afforded by restricting to two spacetime dimensions allow for a detailed study of not only bulk effects but even secondary effects such as backreaction \cite{Almheiri:2014cka,Guo:2022ivd,Gaikwad:2022jar}, the understanding of which is pivotal for any resolution of the black hole information paradox. 
    In the JT gravity side, this backreaction is modeled by coupling in an additional free scalar which modifies the Schwarzian theory and introduces divergences in the moduli-space integral \cite{Moitra:2022glw}. 
    These divergences in turn can be understood by  coupling the dual SYK model to a set of scalars to give a random Yukawa-SYK-like contribution \cite{Moitra:2022glw},
    \bea
      -\sum_{j=1}^M\sum_{1\le j_1<j_2<j_3<j_4\le N} g_{j_1j_2j_3j_4, j}\psi_{j_1}\psi_{j_2}\psi_{j_3}\psi_{j_4}\frac{\phi_j}{\sqrt{M}}\,, 
      \nn\\
    \eea
    to the Hamiltonian. 
    This introduces a new boundary parameter (corresponding to the bulk backreaction parameter) that controls the strength of the backreaction.
\end{itemize}
This last point will form the seed of the current article. 
Quite apart from its connection to quantum gravity, the SYK model has also played a pivotal role in recent developments in the condensed matter physics of strange metals as well as quantum information theory. 
In part, this is because, with its finite-dimensional Hilbert space and emergent large-$N$ properties, it offers a controllable setting to study quantum information measures such as entanglement entropy and quantum chaos. 
With this in mind, the central question that we set out to answer in this letter is: {\it What are the holographic implications of the effect of backreaction, through the lens of quantum entanglement and chaos in the SYK model?}
\\

\noindent
To reiterate, the finite-$N$ SYK model describes a fermionic quantum mechanical system with a finite-dimensional Hilbert space. 
Consequently, its algebra of observables, the set of operators that can measure or manipulate the quantum states in the Hilbert space, belongs to the class of von Neumann algebras of Type I (vN I), a property it is expected to share with a theory of Quantum Gravity. 
The other types of von Neumann algebras are von Neumann Type II (vN II) and Type III (vN III), which are relevant for quantum field theory and pure gravity respectively. 
These algebras can be distinguished from vN I algebras by the fact that they do not allow for the definition of a pure state alone, since they do not contain minimal projections, which are the only operators that can isolate pure states. 
They also have different ways of representing the density matrices \cite{Leutheusser:2021qhd}. 
vN II and vN III algebras can be further classified into subtypes, depending on some additional properties. 
For example, vN II algebras can be either vN II$_{1}$ or vN II$_{\infty}$, depending on whether they contain a maximally mixed state (vN II$_{1}$) or not (vN II$_{\infty}$). 
In essence, vN algebras distinguish {\it physical properties} of different systems and should be general enough for exploring the mechanism of emergence of spacetime. 
In particular, the graviton and bulk matter are simultaneously emergent in holographic models when $N\to\infty$, a fact that should be coded in the van Neumann algebra of observables.
\\

\noindent
In this letter, we derive and analyze the entanglement and chaos spectrum of the SYK model coupled to Dirac fermions. 
Unlike in the case of bosonic matter, the boundary degrees of freedom in this model are always finite ensuring that we do not need to introduce a cutoff when computing physical observables such as the spectrum. 
We find that there is a strong correlation between the graviton and matter fields, leading to {\it maximal entanglement} after a sufficiently long period. 
This is matched by a shift from vN I algebra to vN II algebra as we approach the large-$N$ limit. 
The thermal field double state was used to obtain the analytical results of vN algebra in JT gravity and SYK model \cite{Chandrasekaran:2022qmq,Penington:2023dql}.
Our numerical result provides the dynamical analysis of vN algebra. 
To analyze the spectral statistics of this backreacted SYK system, we also compute the averaged nearest-neighbor gap ratio \cite{Atas:2012prl}. 
The presence of backreaction from Dirac fermions does not seem to prevent the occurrence of a random matrix distribution. 
Our results, based on the classification of the random matrix distribution from charge-conjugation operators as well as the observation of spectral degeneracy, are compatible with, and similar to the SYK model.

\section{Low-Energy Theory and SYK Model}
\label{sec:2} 
\noindent 
By way of reminder, and to establish our conventions, the action for JT gravity  consists of two terms \cite{Teitelboim:1983ux,Jackiw:1984je}; one describing the bulk gravity and the other a Gibbons-Hawking boundary term, 
\bea
   S_{\mathrm{JT}}&=&-\frac{1}{16\pi G_2}\int_{\cal M} d^2x\sqrt{|\det{g_{\mu\nu}}|}\ \phi (R-2\Lambda)
   \nn\\
   &&-\frac{1}{8\pi G_2}\int_{\partial {\cal M}} du \sqrt{|\det{h_{uu}}|}\ \phi K\,. 
\eea   
Here $g_{\mu\nu}$ is the 2-dimensional metric on ${\cal M}$, $\phi$ is the dilaton field, $h_{uu}$ is the induced metric on the boundary, $\partial {\cal M}$, and $R$ is the scalar curvature. The cosmological constant $\Lambda$ introduces a constant curvature term into the spacetime manifold and $G_2$ is the Newton constant in (1+1) dimensions. 
The second, boundary term involves the trace of the extrinsic curvature $K$ of the boundary. 
The fields are subject to Dirichlet boundary conditions, which fix their values on the spacetime boundary. 
To this we add $\widetilde{M}=M/2$ bulk Dirac fermions,
\bea
S_{\mathrm{DF}}=\frac{\lambda}{32\pi G_2\widetilde{M}}\sum_{k=1}^{\widetilde{M}}\int_{{\cal M}} d^2x\sqrt{|\det g_{\rho\sigma}|}\ 
\bar{\Psi}_k\bar{\gamma}^{\mu}\overleftrightarrow{D}_{\mu}\Psi_k,   
\nn\\
\eea
where $\bar{\Psi}\equiv\Psi^{\dagger}\gamma^0$ and $\bar{\gamma}^{\mu}\equiv e_a{}^{\mu}\gamma^a$. 
Latin indices $a,b,\cdots$ are vielbein indices, and $M$ counts the number of {\it boundary} Dirac fermion fields. 
The $\gamma$-matrices in the vielbein basis are given by
 \bea
 \gamma_0=\sigma_x=\begin{pmatrix}
 0&1
 \\
 1&0
 \end{pmatrix}, \ 
 \gamma_1=-\sigma_y=\begin{pmatrix}
 0&i
 \\
 -i&0 
 \end{pmatrix}. 
 \eea
The derivative operator $\overleftrightarrow{D}_{\nu}$ acts as $\bar{\Psi}\bar{\gamma}_{\mu}\overleftrightarrow{D}_{\nu}\Psi=\bar{\Psi}\bar{\gamma}_{\mu}\overrightarrow{D}_{\nu}\Psi-\bar{\Psi}\overleftarrow{D}_{\nu}\bar{\gamma}_{\mu}\Psi$,  
 where 
 \bea
 \overrightarrow{D}_{\mu}&\equiv&\overrightarrow{\partial}_{\mu}+\frac{1}{8}\eta_{ca}\omega_{\mu}{}^c{}_b\lbrack\gamma^a, \gamma^b\rbrack, \ 
 \nn\\
  \overleftarrow{D}_{\mu}&\equiv&\overleftarrow{\partial}_{\mu}-\frac{1}{8}\eta_{ca}\omega_{\mu}{}^c{}_b\lbrack\gamma^a, \gamma^b\rbrack, 
 \eea 
 and the spin connection $\omega_{\mu}{}^a{}_b=-e_b{}^{\nu}(\partial_{\mu}e^a{}_{\nu}-\Gamma^{\lambda}_{\mu\nu}e^a{}_{\lambda})$. 
 Our strategy will be to first derive the boundary theory from the AdS$_2$ JT gravity. 
 This will be identified as a low-energy limit of an SYK model with the coupling term from the bulk Dirac fermion fields informing a corresponding coupling of $M$ boundary Dirac fermions to the SYK model. 

\subsection{Backreaction in the JT gravity} 
\noindent 
We start by integrating out the dilaton. 
Since it enters linearly in the action, this amounts to using its equation of motion $R=2\Lambda$ \cite{Teitelboim:1983ux,Jackiw:1984je}. The spacetime manifold ${\cal M}$ is therefore just a surface of constant curvature, either positive or negative depending on the sign of the cosmological constant. For definiteness, we will consider a negative cosmological constant in this letter. 
The resulting AdS$_2$  metric, 
\bea
   ds^2=-\frac{1}{\Lambda}\frac{dt^2+dz^2}{z^2}\,,
\eea  
can be simplified by using the conformal gauge,
\bea
   ds^2&=&-2e^{2\rho}dx^+dx^-, \qquad e^{2\rho}=-\frac{2}{\Lambda}\frac{1}{(x^+-x^-)^2},
\eea
where $x^{\pm}\equiv t\pm iz$, and the vielbein basis is defined through $g_{\mu\nu}=e_{\mu}{}^ae_{\nu}{}^b\eta_{ab}$. 
The equation of motion of the Dirac fermion field is 
\bea
i\bar{\gamma}_{\mu}\overrightarrow{D}^{\mu}\Psi=0,
\label{EOMF}
\eea
 where $\Psi = (\Psi_{1},\Psi_{2})^{\mathrm{T}}$ is a two-component Dirac spinor. 
Finally, the equation of motion of the metric field is 
\bea
   -\frac{1}{8\pi G_2}(\nabla_{\mu}\nabla_{\nu}\phi
   -g_{\mu\nu}\nabla^2\phi-\Lambda g_{\mu\nu}\phi)=T_{\mu\nu}, 
   \label{EOMM}
\eea
where 
\bea
\frac{1}{2}\int_{{\cal M}} d^2x\sqrt{|\det{g_{\rho\sigma}}}|\ T_{\mu\nu}\equiv \frac{\delta S_{\mathrm{DF}}}{\delta g^{\mu\nu}}. 
\eea 
Note that the energy-momentum tensor is only non-zero for the $\pm\pm$ components. 
Moreover, since integrating the dilaton fixes the metric, any backreaction only affects the dilaton field through Eq. \eqref{EOMM} at the classical level. 
\\

\noindent  
Since the fermionic matter is free, w.l.o.g., we will take $\widetilde{M}=1$ for our derivation of the low-energy theory; the case $\widetilde{M}>1$ yields a similar result. In terms of the individual spinor components, Eq. \eqref{EOMF} can be rewritten as \cite{Guo:2022ivd}:
\bea
  2\partial_-\Psi_1=-\frac{\Psi_1}{x^+-x^-}\,, \quad 
  2\partial_+\Psi_2=\frac{\Psi_2}{x^+-x^-}. 
\eea
These are subsequently easily solved to give: 
\bea
\Psi_1&\sim& (x^+-x^-)^{\frac{1}{2}}\int^{\infty}_{-\infty} d\tau\frac{j_1(\tau)}{|\tau-x_+|}, \quad
\nn\\ 
\Psi_2&\sim& (x^+-x^-)^{\frac{1}{2}}\int^{\infty}_{-\infty} d\tau\frac{j_2(\tau)}{|\tau-x_-|}, 
\eea 
where $\sim$ denotes equivalence up to an overall constant. The boundary action from the backreaction is then:  
\bea
&&
\frac{\lambda}{16\pi G_{2}}\int dt\ \bar{\Psi}\bar{\gamma}_z\Psi
\nn\\
&\sim&
-\frac{\lambda}{32\pi G_2}\int d\tau_1d\tau_2\ \frac{j_1^*(\tau_1)j_1(\tau_2)}{|\tau_1-\tau_2|}
\nn\\
&&
-\frac{\lambda}{32\pi G_2}\int d\tau_1d\tau_2\ \frac{j_2^*(\tau_1)j_2(\tau_2)}{|\tau_1-\tau_2|}
\nn\\
&=&
-\frac{\lambda}{32\pi G_2}\int d\tau_1d\tau_2\ \frac{J^{\dagger}(\tau_1)J(\tau_2)}{|\tau_1-\tau_2|}
\nn\\
&=&
-\frac{\lambda}{32\pi G_2}\int^{\infty}_{-\infty}du_1du_2\ \frac{f^{\prime}(u_1)f^{\prime}(u_2)}{|f(u_1)-f(u_2)|}
\tilde{J}^{\dagger}(u_1) \tilde{J}(u_2).  
\nn\\
\label{LTB}
\eea
Here $J\equiv\begin{pmatrix} j_1& j_2\end{pmatrix}^T$ is a boundary field and $\tilde{J}$ is the same field after a reparametrization of the boundary coordinate. 
This reparametrization does not change the field, $j_k(\tau)=\tilde{j}_k(u)$. 
The Lagrangian that encodes the backreaction is invariant under the same SL(2) transformation as the Schwarizan theory 
\cite{Jensen:2016pah,Maldacena:2016upp,Engelsoy:2016xyb},
\bea
   S_{\mathrm{SCH}}=-\frac{1}{8\pi G_2}\int^{\infty}_{-\infty}du\ 
   \phi_b\bigg(\frac{f^{\prime\prime\prime}}{f^{\prime}}-\frac{3}{2}\frac{f^{\prime\prime 2}}{f^{\prime 2}}\bigg)\,,  
\eea 
that is the boundary theory for JT gravity. Following convention in Refs. \cite{Jensen:2016pah,Maldacena:2016upp,Engelsoy:2016xyb}, and taking $\epsilon$ as a boundary cutoff that fixes the length of the boundary so that 
\bea
   \frac{1}{|\Lambda|}\frac{t^{\prime 2}(u)+z^{\prime 2}(u)}{z^2}=\frac{1}{\epsilon^2}\,, 
\eea 
the bulk dilaton field approaches the boundary as 
\bea
\lim_{z\rightarrow 0}\phi=\frac{\sqrt{|\Lambda|}}{\epsilon}\phi_b\,. 
\eea
Hence the boundary theory of the backreacted geometry is the combination of Eq. \eqref{LTB} and the Schwarzian theory. 
We can transform the boundary time coordinate so that it has the same asymptotic conditions as the dilaton without the backreaction \cite{Almheiri:2014cka}. 
\\

\noindent
The promotion of the reparametrization {\it function} $f$ to a boundary {\it field} requires that the solution space of the bulk and boundary theories be equivalent. 
To establish this, let us examine the equation of motion for $f$ which reads,
\bea
\bigg\lbrack
\frac{1}{f^{\prime}(u)}
\bigg(\frac{\big(f^{\prime}(u)\phi_b(u)\big)^{\prime}}{f^{\prime}(u)}
\bigg)^{\prime}
\bigg\rbrack^{\prime}
=B(u)\,.   
\eea
Here $B(u)$ is a general function of $u$ and denotes the source of the backreaction. Solving this for the boundary field gives, at least formally,
\bea
\phi_b(u)=\frac{c_1+c_2f(u)+c_3f^2(u)}{f^{\prime}(u)}\,, 
\eea
where $c_1$, $c_2$, and $c_3$ are arbitrary constants into which the backreaction effects can be absorbed. The backreaction does not generate a new solution space. 
We conclude that the SL(2)-invariant path integration measure should provide a proper boundary description in a similar manner as in Refs. \cite{Jensen:2016pah,Maldacena:2016upp,Engelsoy:2016xyb}. 

\subsection{Backreaction in SYK Model} 
\noindent
The Hamiltonian for the $\mathrm{SYK}_q$ model describes the quantum dynamics of $N$ Majorana fermions with random $q$-body interactions. 
The Hamiltonian can be written as
\begin{eqnarray}
   H_{\mathrm{SYK}_{q}} = -i^{\frac{q}{2}} \sum_{1 \le i_1 < i_2 < \cdots < i_q \le N} J_{i_1 i_2 \cdots i_q} \psi_{i_1} \psi_{i_2} \cdots \psi_{i_q}\,,
   \nn\\
\end{eqnarray}
where $\psi_i$ are the Majorana fermion operators that satisfy the anticommutation relations $\{\psi_i, \psi_j\} = \delta_{ij}$, and $J_{i_1 i_2 \cdots i_q}$ are random coupling constants that follow a Gaussian distribution with zero mean and variance,
\begin{eqnarray}\label{GD}
   \overline{J_{i_1 i_2 \cdots i_q}^2} = \frac{(q-1)! J^2}{N^{q-1}}\,,
\end{eqnarray}
where $J$ is a constant that sets the energy scale of the model. 
The Hamiltonian is invariant under the O($N$) symmetry that rotates the Majorana fermions. The parameter $q$ determines the number of fermions that interact in each term of the Hamiltonian and can be any even integer greater than or equal to two.
\\

\noindent 
Following the strategy outlined above, and utilizing the two-point correlator of the SYK model,
\bea
\frac{1}{N}\sum_{j=1}^N\langle \psi_j(t_1)\psi_j(t_2)\rangle\sim\frac{\sgn(\tau_1-\tau_2)}{|t_1-t_2|^{\frac{1}{2}}}\,, 
\eea 
it is not difficult to see that the coupling of an SYK$_2$ term and a Dirac fermion results in the low-energy effective action \eqref{LTB}. 
By integrating out the random coupling constant, one can reach the point where four Majorana fermions interact with each other. 
More generally, coupling in $M$ boundary Dirac fermions, the ($q=4$) Yukawa-like SYK Hamiltonian that will be the focus of the rest of this article is given by,
\bea
H_{\mathrm{YSYK}}
&=&
\sum_{1\le i_1<i_2<i_3<i_4\le N} j_{i_1i_2i_3i_4}\psi_{i_1}\psi_{i_2}\psi_{i_3}\psi_{i_4}
\nn\\
&&
+\frac{i}{\sqrt{M}}\sum_{j=1}^{M}\sum_{1\le i_1<i_2\le N} (g_{i_1i_2, j}\psi_{i_1}\psi_{i_2}
\Psi_{j}^{\dagger}
\nn\\
&&
+g_{i_1i_2, j}^*\psi_{i_1}\psi_{i_2}
\Psi_{j})\,, 
\label{HSYK}
\eea
where now, the random couplings are drawn from the distribution
\bea
&&
\exp\Bigg(-\sum_{1\le i_1<i_2<i_3<i_4\le N}j_{i_1i_2i_3i_4}^2\frac{N^{3}}{12J^2}\Bigg)
\nn\\
&&
\times
\exp\Bigg(-\sum_{j=1}^M\sum_{1\le i_1<i_2\le N}g_{i_1i_2, j}^*g_{i_1i_2, j}\frac{N}{2g^2}\Bigg)\,, 
\eea 
and $g$ controls the strength of the backreaction. 
Since the bulk Dirac fermion is a two-component spinor, the number of (one-component) boundary fermion fields $M$ must be even. Taking $\beta$ to be the inverse temperature, the gravitational theory emerges in the following holographic limit: 
\bea
G_2\sim \frac{1}{N}\ll 1; \ \phi_b\sim\frac{1}{\beta J}\ll 1; \ \lambda\sim\frac{g^2}{J^2}\ll 1\,. 
\eea 
What follows next is a detailed numerical study of the entanglement and spectral properties of the Hamiltonian \eqref{HSYK}.

\section{Entanglement Entropy and von Neumann Algebras}
\label{sec:3}
\begin{figure}
\begin{center}
    \includegraphics[width=6.0cm]{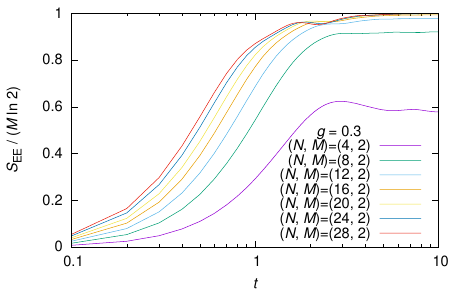} \includegraphics[width=6.0cm]{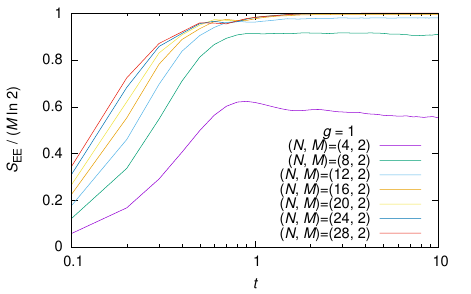}
    \includegraphics[width=6.0cm]{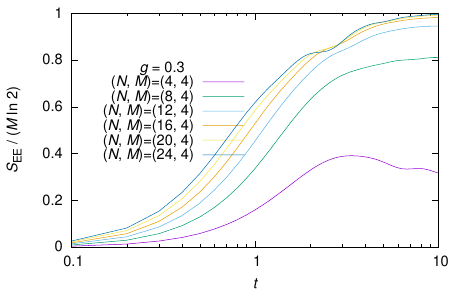} \includegraphics[width=6.0cm]{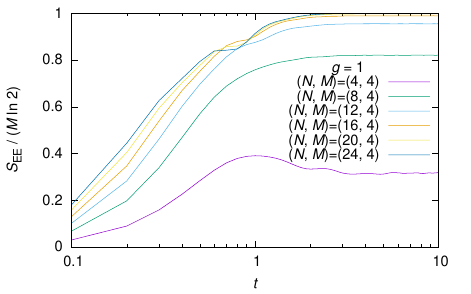}
    \end{center}
    \begin{center}    \includegraphics[width=6.0cm]{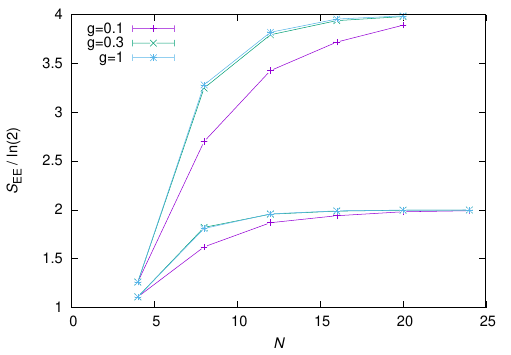}  \end{center}
    \caption{
    [Top and Middle] Entanglement entropy for [top, left to right] $M=2$ and $g=0.3$ and $1$, [second row] $M=4$ and $g=0.3$ and $1$. 
    [Bottom] Late-time value of the entanglement entropy as a function of $N$ for $M=4$ (upper) and $M=2$ (lower), computed as the average over $t=(1,2,\cdots,100)\times10^3$. 
    When $S_{\mathrm{EE}}/\big(M\ln(2)\big)$ reaches 1, the Majorana and matter fields are maximally entangled.
    }
    \label{fig:EntanglementEntropyN=2M}
\end{figure}

\noindent 
We begin our numerical investigation of the entanglement entropy of the boundary theory \eqref{HSYK} as usual by partitioning the system into two subsystems, say $A$ and $B$ and computing $S_{\mathrm{EE}}=-\mathrm{Tr}_A\left(\rho_A\ln\rho_A\right)$, where $\mathrm{Tr}_A$ and $\rho_A$ are the partial trace operation and the reduced density matrix on the region $A$, respectively. 
In our case, the Hilbert space is $\mathscr{H}_M\otimes \mathscr{H}_D$, where $\mathscr{H}_M$ and $\mathscr{H}_D$ are the Hilbert spaces of the (Majorana) SYK model and (Dirac) matter sector, respectively. 
We take the partial trace over the matter sector to obtain the entanglement entropy with a fixed initial state, $|0\cdots 0\rangle$, which is the unoccupied state in the occupation number basis. 
Our simulations suggest that ({\it i}) in general, the entanglement entropy first increases and then saturates, as shown in Fig.~\ref{fig:EntanglementEntropyN=2M}; ({\it ii}) it saturates to the maximally allowed value of $M\ln 2$ in the large-$N$ limit and; ({\it iii}) that the saturation time is finite in the large-$N$ limit.
From these numerical observations, we deduce then that the state is well approximated by one that is maximally entangled between the Majorana fermion and Dirac matter fields. 
Using the Schmidt decomposition theorem, one can find a pair of orthonormal bases to express the state in the following form 
\bea
   |\psi\rangle=\frac{1}{\sqrt{2^M}}\sum_{j=1}^{2^M}|j_{M}\rangle\otimes|j_D\rangle\,, 
\eea
when reaching the saturation regime in the large-$N$ limit. Here $|j_M\rangle$ and $|j_D\rangle$ are some orthonormal bases for $\mathscr{H}_M$ and $\mathscr{H}_D$, respectively. 
The saturation regime exhibits a transition in its algebra of observables from a vN I algebra to a vN II algebra as $N\rightarrow\infty$ implying that the graviton is maximally entangled with the Dirac fermion matter fields from the bulk perspective. 
Having deduced a plausible state, we can now take the partial trace over the matter fields to extract the von Neumann algebra of the graviton. 
The maximally mixed state shows that the Majorana fermion algebra is type II$_1$ when $N=2M\rightarrow\infty$. 
This subalgebra is not relevant to the other region that we partial-trace over. 
In the saturation regime, however, the algebra of the large-$N$ theory is equivalent to a vN II$_1$ algebra. 
We therefore identify an algebraic transition from vN I to vN II, which is in turn suggestive of a loss of information.  

\section{Random Matrix Theory}
\label{sec:4}
\noindent 
The Hamiltonian \eqref{HSYK} contains an odd number of fermion couplings and consequently does not commute with the particle-hole operators
\bea
{\cal H}\equiv K\prod_{j=1}^{T_d}(c_j+c^{\dagger}_j)\,, 
\eea
where $T_d\equiv N/2+M$, $K$ is an anti-linear operator and acts as complex conjugation, and the Dirac operators ($c_j$ and $c_j^{\dagger}$) remain real. 
The first term in the Hamiltonian \eqref{HSYK} involves the product of four Majorana fermions, while the second involves the product of two Majorana fermions. 
This implies that the number parity of complex fermions created from these Majorana fermions remains conserved, and the matrix representation of Eq. \eqref{HSYK} is block-diagonal. In other words, the parity operator is almost the same as the SYK model but acts trivially on the matter sector.
\\

\begin{figure}
\centering
    \includegraphics[width=6.0cm]{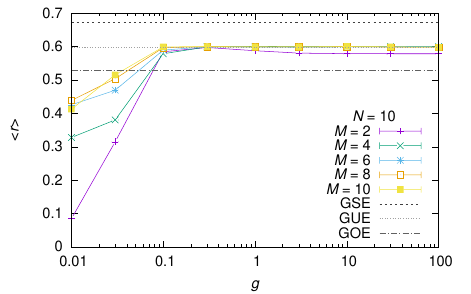}
    \includegraphics[width=6.0cm]{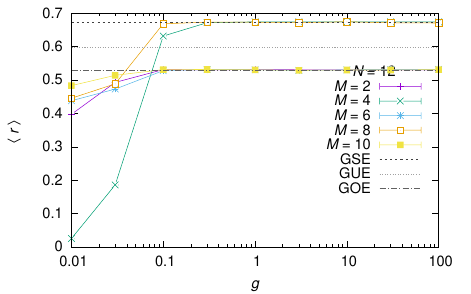}\\
    \includegraphics[width=6.0cm]{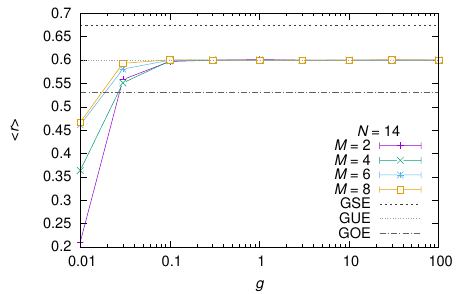}
    \includegraphics[width=6.0cm]{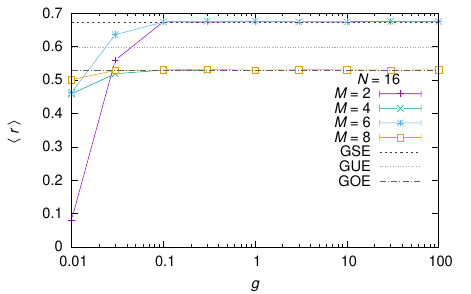}
    \caption{The average of neighboring gap ratio $\langle r\rangle$, where the average over all neighboring gap pairs (see Eq.~\eqref{eqn:r}) is taken, plotted against $g$.
    For $N=12$, $M=2, 4, 6, 8, 10$.
    For $N=14$, $M=2, 4, 6, 8$.
    For $N=16$, $M=2, 4, 6$. 
    For comparison, we have also included the values from the random matrix theory \cite{Atas:2012prl} in the plot.}
    \label{fig:RandomMatrixTheory}
\end{figure}

\noindent
To explore the spectral statistics of the model, we first identify the nondegenerate eigenvalues $\{E_1,E_2,\ldots,E_D\}$ in each parity sector, compute the gaps $\{\delta_1=E_2-E_1, \delta_2=E_3-E_2, \ldots, \delta_{D-1}=E_D-E_{D-1}\}$, 
and the spectral gap ratio
\bea
\{r_j\}_{j=1}^{D-2}=\left\{\frac{\min(\delta_j,\delta_{j+1})}{\max(\delta_j,\delta_{j+1})}\right\}_{j=1}^{D-2}.
\label{eqn:r}
\eea
As is well-known, this ratio can distinguish between different types of random matrices, such as the Gaussian orthogonal ensemble (GOE), the Gaussian unitary ensemble (GUE), and the Gaussian symplectic ensemble (GSE), that are characterized by different symmetries of the random matrices. 
For large Gaussian random matrix ensembles, the average of $r_j$ is known to take values
0.5307(1) for GOEs, 
0.5996(1) for GUEs, and 
0.6744(1) for GSEs \cite{Atas:2012prl}. 
Written in terms of the Majorana operators $\gamma_{j}$, it is easy to see that the charge-conjugation operators,
\bea
{\cal P}\equiv K\prod_{j=1}^{T_d}\gamma_{2j-1}; \ {\cal R}\equiv K\prod_{j=1}^{T_d}i\gamma_{2j} 
\label{cc}
\eea
still commute with the Hamiltonian and, as a result, continue to generate the symmetries of the Hamiltonian relevant to classifying the random matrix distribution. 
Our numerical observations of the degeneracy of the spectrum and the approximate description of random matrix properties of the boundary theory are captured in Table~\ref{tbl:SpectralDegeneracy} and Fig.~\ref{fig:RandomMatrixTheory} respectively.
\begin{table}
    \caption{Degeneracy in the spectrum. 
    When we compare our model with the SYK model having $2T_d$ fermions, we observe the same degeneracy within each parity sector. 
    }
    \label{tbl:SpectralDegeneracy}
    \centering
    \begin{tabular}{|c||c|c|c|c|c|c|c|c|}
    \hline $N$ \textbackslash $M$ & 2 & 4 & 6 & 8 & 10 & 12 \\\hline\hline
    4                      & 1 & 4 &16 &64 &256 & 1024 \\\hline
    6                      & 1 & 1 & 1 & 1  &  4 &  16  \\\hline
    8                      & 2 & 1 & 2 & 1 &  2 &  1 \\\hline
    10                     & 1 & 1 & 1 & 1 &  1  &  1 \\\hline
    12                     & 1 & 2 & 1 & 2 &  1   &    \\\hline
    14                     & 1 & 1 & 1 & 1 &        &    \\\hline
    16                     & 2 & 1 & 2 & 1  &        &    \\\hline
    18                     & 1 & 1 & 1 &    &        &    \\\hline
    20                     & 1 & 2 & 1 &     &   &            \\\hline
    22                     & 1 & 1 &   &      &   &            \\\hline
    24                     & 2 & 1 &   &      &   &              \\\hline
    \end{tabular}
\end{table}
The degeneracy patterns and the random matrix properties for sufficiently large $N/M$ can be summarized as follows,
 \begin{itemize}
\item{$T_d \mod 4 =0, ({\cal P}^2, {\cal R}^2)=(1, 1)$, no degeneracy in the spectrum, GOE; 
}
\item{$T_d \mod 4 =1, ({\cal P}^2, {\cal R}^2)=(1, -1)$, no degeneracy in the spectrum, GUE; 
}
\item{$T_d \mod 4 =2, ({\cal P}^2, {\cal R}^2)=(-1, -1)$, twofold degeneracy in the spectrum, GSE; 
}
\item{$T_d \mod 4 =3, ({\cal P}^2, {\cal R}^2)=(-1, 1)$, no degeneracy in the spectrum, GUE.   
}
\end{itemize} 
However, to properly analyze the distribution of random matrices, a thorough treatment of $1/N$ corrections will be necessary. 
Nevertheless, it is evident that the backreaction does not prevent the emergence of random matrix behavior at large $N$. 
\\

\noindent
From the results presented above, we conclude that quantum information measures are useful tools for analyzing holographic theory, particularly in the presence of Dirac fermion backreaction. 
The backreaction of Dirac fermion fields introduces new features to the JT gravity theory, such as maximal entanglement, algebraic transition, and 3-fermion coupling. 
We anticipate that the 3-fermion coupling in particular may have a role to play in the physics of topological quantum matter. 
On the other hand, the simplicity of the model also makes it a useful tool for a deeper exploration of the mechanisms underlying the emergence of spacetime.

\section*{Acknowledgments}
\noindent 
PHCL acknowledges the support from JSPS KAKENHI (Grant No. 20H01902 and JP23H01174), MEXT KAKENHI (Grant No. 21H05462). 
CTM acknowledges the Nuclear Physics Quantum Horizons program through the Early Career Award (Grant No. DE-SC0021892); 
YST Program of the APCTP. 
JM would like to acknowledge support from the ICTP through the Associates Programme, from the Simons Foundation (Grant No. 284558FY19), and from the ``Quantum Technologies for Sustainable Development''  grant from the National Institute for Theoretical and Computational Sciences of South Africa (NITHECS).
MT acknowledges the Grants-in-Aid from MEXT of Japan (Grant No. JP20H05270, No. JP20K03787, and No. JP21H05185).
CTM thanks Nan-Peng Ma for his encouragement. 
We thank the Shanghai University; 
Center for Quantum Spacetime, Sogang University; 
National Chengchi University; 
National Tsing Hua University; 
Great Bay University, 
where we presented this work. 



\end{document}